\documentclass[10pt, conference, letterpaper]{IEEEtran}

\RequirePackage{graphics}
\usepackage{epsfig}
\usepackage{amsmath}
\usepackage{amsfonts}
\usepackage{graphicx,subfigure}
\usepackage{amssymb}
\usepackage{algorithm}
\usepackage{algorithmic}
\usepackage{color}
\usepackage{framed}
\usepackage{xcolor}
\usepackage{lipsum}

\def\overbracket#1{\mathop{\vbox{\ialign{##\crcr\noalign{\kern3\p@}
\downbracketfill\crcr\noalign{\kern3\p@\nointerlineskip}
$\hfil\displaystyle{#1}\hfil$\crcr}}}\limits}
\def\underbracket#1{\mathop{\vtop{\ialign{##\crcr
$\hfil\displaystyle{#1}\hfil$\crcr\noalign{\kern3\p@\nointerlineskip}
\upbracketfill\crcr\noalign{\kern3\p@}}}}\limits}

\usepackage{cite}

\ifCLASSINFOpdf
\else
\fi

\begin{document}
\title{Hiding Communications in AWGN Channels and THz Band with Interference Uncertainty}

\author{\IEEEauthorblockN{
Zhihong Liu\IEEEauthorrefmark{1}, Jiajia
Liu\IEEEauthorrefmark{2}\IEEEauthorrefmark{5}, Yong Zeng\IEEEauthorrefmark{1},
and Jianfeng Ma\IEEEauthorrefmark{1}}

\IEEEauthorblockA{\IEEEauthorrefmark{1}School of Cyber Engineering, Xidian University, China}

\IEEEauthorblockA{\IEEEauthorrefmark{2}School of Cyberspace Security, Northwest Polytechnic University, China}

\IEEEauthorblockA{\IEEEauthorrefmark{5}E-mail: liujiajia@nwpu.edu.cn} }


\maketitle

\begin{abstract}
Covert communication can prevent an adversary from knowing that a wireless transmission has occurred. In
additive white Gaussian noise (AWGN) channels, a square root law is found that  Alice can reliably and covertly
transmit $\mathcal{O}(\sqrt{n})$ bits to Bob in $n$ channel uses. In this paper, we consider covert
communications in noisy wireless networks, where the receivers not only experience the background noise, but also
the aggregate interference from other transmitters. Our results show that uncertainty in interference
experienced by the adversary Willie is beneficial to Alice. In AWGN channels, when the distance between Alice
and Willie $d_{a,w}=\omega(n^{1/(2\alpha)})$ ($\alpha$ is the path loss exponent), Alice can reliably and
covertly transmit $\mathcal{O}(\log_2\sqrt{n})$ bits to Bob in $n$ channel uses. Although the covert
throughput is lower than the square root law, the spatial throughput is higher. In THz (Terahertz) Band networks,
covert communication is more difficult because Willie can simply place a receiver in the narrow beam between
Alice and Bob to detect or block their LOS (Line-of-Sight) communications. We then present a covert
communication scheme that utilizes the reflection or diffuse scattering from a rough surface to prevent being
detected by Willie. From the network perspective, the communications are hidden in the interference of noisy
wireless networks, and what Willie sees is merely a ``shadow'' wireless network.
\end{abstract}

\begin{IEEEkeywords}
Physical-layer Security; Covert Communications; AWGN Channel; THz Band.
\end{IEEEkeywords}

\IEEEpeerreviewmaketitle

\section{Introduction}
Traditional cryptography methods for network security can not solve all security problems. In wireless networks,
if a user wishes to communicate covertly (without being detected by other detectors), encryption to preventing
eavesdropping is not enough \cite{Hiding_Information}. Even if a message is encrypted, the metadata, such as
network traffic pattern, can reveal some sensitive information \cite{ICCCN}. Furthermore, if the adversary
cannot detect the transmissions, he has no chance to launch the ``eavesdropping and decoding'' attack even if he
has boundless computing and storage capabilities. In a battlefield, soldiers hope to hide their tracks so they need
to communicate covertly. Another occasion, such as defeating ``Panda-Hunter'' attack \cite{panda_hunter}, also
needs to prevent the adversary from detecting transmission behavior to protect user's location privacy.

Consider a scenario that a transmitter Alice would like to send a message to a receiver Bob covertly over a
wireless channel in order to not being detected by a warden Willie. In \cite{square_law}, Bash \emph{et al.} found
a square root law in additive white Gaussian noise (AWGN) channels, that is, Alice can transmit
$\mathcal{O}(\sqrt{n})$ bits reliably and covertly to Bob over $n$ uses of wireless channels. The square root
law implies pessimistically that the asymptotic privacy rate approaches zero. If Willie does not know the time of
the transmission attempts of Alice, Alice can reliably transmit $\mathcal{O}(\min\{(n\log T(n))^{1/2}, n\})$ bits
to Bob while keeping Willie's detector ineffective with a slotted AWGN channel containing $T(n)$ slots
\cite{time1}. To improve the performance of covert communication, Lee \emph{et al.} \cite{LDP1} found that,
Willie has measurement uncertainty about its noise level due to the existence of SNR wall \cite{SNR}, then they
obtained an asymptotic privacy rate which approaches a non-zero constant. Following Lee's work, He \emph{et
al.} \cite{Biao_He} defined new metrics to gauge covertness of communication, and Liu \emph{et al.}
\cite{ICC2018} took the interference measurement uncertainty into considerations. Wang \emph{et al.}
\cite{Fundamental_Limits} considered covert communication over discrete memoryless channels (DMC), and found
that privacy rate scales like the square root of blocklength. Bloch \emph{et al.} \cite{Bloch} discussed the covert
communication problem from a resolvability perspective. He developed an alternative coding scheme such that, if
warden's channel statistics are known, on the order of $\sqrt{n}$ reliable covert bits may be transmitted to Bob
over $n$ channel uses with only on the order of $\sqrt{n}$ bits of secret key.

In general, the covertness of communication is due to the
existence of noise that the adversary cannot accurately distinguish between signal and noise.
If we can increase the measurement uncertainty of adversary, the performance of covert
communication can be improved.

Interference or jamming is usually considered harmful to wireless communications, but it is also a useful security
tool. Cooperative jamming is regarded as a prevalent physical-layer security approach
\cite{Physical-Layer-Security}\cite{Challenges}\cite{heartbeats}. Sobers \emph{et al.}
\cite{jammer}\cite{jammer1} utilized cooperative jamming to carry out covert communications. To achieve the
transmission of $\mathcal{O}(n)$ bits covertly to Bob over $n$ uses of channel, they added a ``jammer'' to the
environment to help Alice for security objectives. Liu \emph{et al.} \cite{WCM2018} exploited the interference
from other transmitters in the network to hide the transmission of sensitive information in IoT. Soltani \emph{et
al.} \cite{jammer2}\cite{07096} considered a network scenario where multiple ``friendly'' nodes  generate
interference to hide the transmission from multiple adversaries.

\begin{figure}
\centering \epsfig{file=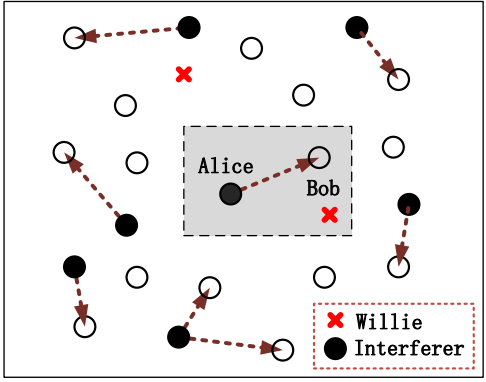, height=2in} \caption{System configuration of covert communication in a wireless network.}\label{deployment}
\end{figure}

In this work, we consider covert communication in a wireless network, where Bob and Willie not only experience
the background noise, but also interference from other transmitters (Fig.\ref{deployment}). Since the measure
uncertainty of aggregate interference is greater than the background noise, the uncertainty of Willie will increase
along with the increase of interference. We consider two kind of communication channels: AWGN channels and
THz (Terahertz) Band channels.  THz Band signals are often assumed to be more secure than lower frequency
signals due to the more directional transmission and the more narrow beams. However this also makes covert
communication difficult. Willie can simply place a receiver in the LOS (Line-of-Sight) path between Tx and Rx to
find and block their communications. Alice and Bob may need resorting to the aggregate interference and NLOS
(Non-Line-of-Sight) communications to improve the security and hiding.

\begin{figure}
\centering \epsfig{file=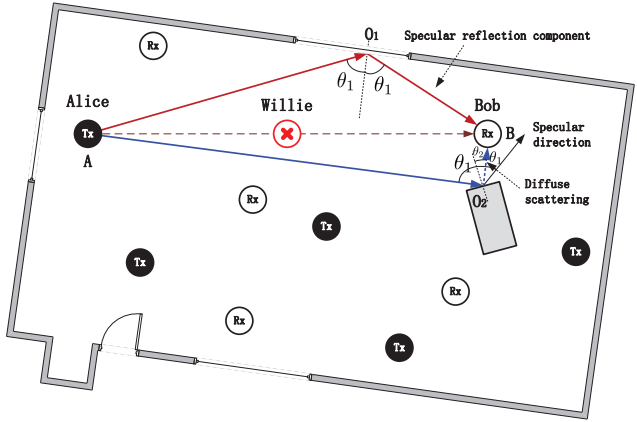, height=2.2in}
\caption{Schematic of covert communication between Alice and Bob with a warden Willie in a THz Band network.
The red lines $\overline{AO_1}$ and $\overline{O_1B}$ denote the specular reflection path,
the blue lines $\overline{AO_2}$ and $\overline{O_2B}$ represent the diffuse scattering path.
The black circles Tx are transmitters, and the white circles Rx are receivers. }\label{thz_model}
\end{figure}

In a dense wireless network with AWGN channels, we found that covert communication between Alice and Bob is
still possible. Alice can reliably and covertly transmit $\mathcal{O}(\log_2\sqrt{n})$ bits in $n$ channel uses
when the distance between Alice and Willie $d_{a,w}=\omega(n^{1/(2\alpha)})$ ($\alpha$ is path loss exponent).
Although the covert throughput is lower than the square root law, the spatial throughput is higher, and Alice does
not presuppose the location knowledge of Willie. In THz Band communication networks, although the LOS
communications can be detected easily by Willie, we found that communication based on reflection or diffuse
scattering may be a possible information hiding method. As depicted in Fig. \ref{thz_model}, the communication
via specular reflection path $\overrightarrow{AO_1B}$ or diffuse scattering path $\overrightarrow{AO_2B}$
can evade detection by Willie. In a dense network, the scattering signals Willie eavesdropping are masked by the
ambient noise and the aggregate interference. From the perspective of network, the noisy wireless channels make
the network ``shadow'' to Willie.

%

\section{Problem Formulation and System Model}\label{ch_2}
In this section, prior to presenting the system model, we give a running example to illustrate the
problem of covert wireless communications discussed in this paper.

\subsection{Motivating Scenario}
Covert communication has a very long history. It is always related with steganography \cite{Steganography} which
conceals messages in audio, visual or textual content. However, steganography is an application layer technique
and is not suitable in physical-layer covert communication. The well-known physical-layer covert communication is
spread spectrum which is using to protect wireless communication from jamming and eavesdropping
\cite{Spread_Spectrum}. Another kind of covert communications is network covert channels
\cite{covert_channel_1}\cite{covert_channel_2} in computer networks. While steganography requires some
contents as covers, a network covert channel requires network protocols as carrier. In this paper, we consider
physical-layer covert communication that employs the background noise and the aggregate interference in
wireless channels to hide user's transmission attempts.

Let us take a source-location privacy problem as an example. In Panda-Hunter Game \cite{panda_hunter}, a
sensor network with a large number of sensors is deployed to monitor the habitat of pandas. As soon as a panda
is  observed by a sensor, this sensor reports the readings to a sink via a multi-hop wireless channel. However, a
hunter (Willie) is wandering around in the habitat in order to capture pandas. The hunter does not care about the
readings of sensors, what he really cares about is the location of the message originator, the first sensor who
discovers the panda. To find the location of pandas, he listens to a sensor in his vicinity to determine whether this
sensor is transmitting or not. If he finds a transmitter, he then searches for the next sensor who is
communicating with this transmitter. Via this method, he can trace back to the message originator and catch the
panda. As a result, the source-location information becomes critical and must be protected  in this occasion.

To tackle this problem, Kamat \emph{et al.} proposed phantom routing techniques to provide source-location
privacy from the perspective of network routing \cite{panda_hunter}. From another point of view, physical-layer
covert communication can provide another kind of solution to Panda-Hunter game. If we can hide the transmission
in noise and interference of noisy wireless channels, the hunter will not be able to ascertain which sensor is
transmitting, and therefore cannot trace back to the source.

\subsection{AWGN Channels}
\subsubsection{Channel Model} Consider a wireless communication scene where Alice (A) wishes to send a
message to Bob (B). Right next to them, a warden Willie (W) is eavesdropping over wireless
channels and trying to find whether or not Alice is transmitting.

We adopt the wireless channel model similar to \cite{square_law}\cite{07096}. Consider a time-slotted system
where the time is divided into successive slots with equal duration. All wireless channels are assumed to suffer
from discrete-time AWGN with real-valued symbols. Alice transmits $n$ real-valued symbols
$\{s_i^{(a)}\}_{i=1}^n$. Bob observes a vector $\{y_i^{(b)}\}_{i=1}^n$, where $y^{(b)}_i = s_i^{(a)}+z^{(b)}_i$,
and $z^{(b)}_i$ is the noise Bob experiences which can be expressed as $z^{(b)}_i = z^{(b)}_{i,0} + I^{(b)}_i$,
where $\{z^{(b)}_{i,0}\}_{i=1}^n$ are independent and identically distributed (i.i.d.) random variables (RVs)
representing the background noise of Bob with $z^{(b)}_{i,0}\sim \mathcal{N}(0, \sigma^2_{b,0})$, and
$\{I^{(b)}_i\}^n_{i=1}$ are i.i.d. RVs characterizing the aggregate interference Bob experiences.

Willie observes a vector $\{y_i^{(w)}\}_{i=1}^n$, where $y^{(w)}_i = s_i^{(a)}+z^{(w)}_i$, and $z^{(w)}_i$ is the
noise Willie experiences which can be expressed as $z^{(w)}_i = z^{(w)}_{i,0} + I^{(w)}_i$, where
$\{z^{(w)}_{i,0}\}_{i=1}^n$ are i.i.d. RVs representing the background noise of Willie with $z^{(w)}_{i,0}\sim
\mathcal{N}(0, \sigma^2_{w,0})$, and $\{I^{(w)}_i\}^n_{i=1}$ are i.i.d. RVs characterizing the aggregate
interference Willie experiences.

Suppose each node in the network is equipped with one omnidirectional antenna. The wireless channel is modeled
by large-scale fading with path loss exponent $\alpha$ ($\alpha \geq 2$). The channel gain $\mathbf{h}_{i,j}$
of channel from $i$ to $j$ is static over the signaling period, all links experience unit mean Rayleigh fading, and
$\sigma^2_{b,0} = \sigma^2_{w,0}$.

\subsubsection{Network Model}
Consider a large-scale wireless network, where the locations of transmitters form a stationary Poisson point
process (PPP)\cite{Haenggi_PPP} $\Pi=\{X_i\}$ on the plane $\mathbb{R}^2$. The density of the PPP is
represented by $\lambda$, denoting the average number of transmitters per unit area. Suppose each potential
transmitter has an associated receiver, the transmission decisions are made independently across transmitters
and independent of their locations for each transmitter, and the transmission power employed for each node are
constant power \footnote{Any other channel models with power control or threshold scheduling will have similar
results with some scale factors.} $P_t$. Let the Euclidean distance between node $i$ and node $j$ is denoted as
$d_{i,j}$. The aggregate interference seen by Bob and Willie are the functional of the underlying PPP
$\Pi=\{X_i\}$ and the channel gain,
\begin{eqnarray}
  I_i^{(b)}  &\equiv & \sum_{k\in\Pi}\sqrt{\frac{P_t}{d_{b,k}^{\alpha}}} \mathbf{h}_{b,k}\cdot s_i^{(k)}\sim \mathcal{N}(0,\sigma^2_{I_b}) \label{eq_1}\\
  I_i^{(w)}  &\equiv & \sum_{k\in\Pi}\sqrt{\frac{P_t}{d_{w,k}^{\alpha}}} \mathbf{h}_{w,k}\cdot s_i^{(k)}\sim \mathcal{N}(0,\sigma^2_{I_w}) \label{eq_2}
\end{eqnarray}
where each $s_i^{(k)}$ is a Gaussian random variable $\mathcal{N}(0,1)$ which represents the signal
of $k$-th transmitter in $i$-th channel use, and
\begin{eqnarray}
\sigma^2_{I_b}&=&\sum_{k\in\Pi}\frac{P_t}{d_{b,k}^{\alpha}}|\mathbf{h}_{b,k}|^2=\sum_{k\in\Pi}\frac{P_t}{d_{b,k}^{\alpha}}\Psi_{b,k} \\
\sigma^2_{I_w}&=&\sum_{k\in\Pi}\frac{P_t}{d_{w,k}^{\alpha}}|\mathbf{h}_{w,k}|^2=\sum_{k\in\Pi}\frac{P_t}{d_{w,k}^{\alpha}}\Psi_{w,k}  \label{eq_4_4}
\end{eqnarray}
are shot noise (SN) processes, representing the powers of the interference that Bob and Willie experience. The
Rayleigh fading assumption implies $\Psi_{i,j}=|\mathbf{h}_{i,j}|^2$ is exponentially distributed with the mean is
$\mathbb{E}[\Psi_{i,j}] = 1$.

The powers of aggregate interferences, $\sigma^2_{I_w}$ and $\sigma^2_{I_b}$, are RVs which are
determined by the randomness of the underlying PPP of transmitters and the fading of wireless channels.
Therefore they are difficult to be predicted. Besides, the closed-form distribution of the aggregate interference
is hard to obtain and we have to bound it. We do not consider the issue of signal detection at Bob, and assume
that  Bob knows when Alice transmits. This can be realized in practice by sharing a secret time-table between
Alice and Bob.

\subsubsection{Hypothesis Testing}
To find whether or not Alice is transmitting, Willie has to distinguish between the following two hypotheses
\begin{eqnarray}
\mathbf{H_0}&:& y_i^{(w)}  =  I_i^{(w)} + z_{i,0}^{(w)} \\
\mathbf{H_1}&:& y_i^{(w)}  =  \sqrt{\frac{P_t}{d_{a,w}^\alpha}}\mathbf{h}_{a,w}\cdot s_i + I_i^{(w)}
+ z_{i,0}^{(w)}
\end{eqnarray}
based on the received vector $\mathbf{y}=\{y_i^{(w)}\}_{i=1}^n$. We assume that Willie employs a radiometer
as his detector, and does the following statistic test
\begin{equation}
    T(\mathbf{y})=\frac{1}{n}\mathbf{y}^H\mathbf{y}=\frac{1}{n}\sum^n_{k=1}y_k^{(w)}*y_k^{(w)}>\gamma
\end{equation}
where $\gamma$ denotes Willie's detection threshold and $n$ is the number of samples.

Let $\mathbb{P}_{FA}$ and $\mathbb{P}_{MD}$ be the probability of false alarm and missed detection. Willie
wishes to minimize his probability of error $\mathbb{P}_e^{(w)} = (\mathbb{P}_{FA}+\mathbb{P}_{MD})/2$, but
Alice's objective is to guarantee that the average probability of error $\mathbf{E}[\mathbb{P}_e^{(w)}] =
\mathbf{E}[\mathbb{P}_{FA}+\mathbb{P}_{MD}]/2>1/2-\epsilon$ for an arbitrarily small positive $\epsilon$.

%

\subsection{THz Band Networks}
Next we briefly look into the THz Band channel model \cite{SINR}, network and blocking model, and rough
surface scattering theory.

\subsubsection{Channel Model}
Suppose each user in the THz Band network is equipped with a directional antenna, and the antenna radiation
pattern is the cone model, i.e., a single cone-shaped beam, whose width determines the antenna directivity.  The
antenna gain $G$ for the main lobe is given by
\begin{equation}\label{q1q}
    G = \frac{2}{1-\cos (\phi/2)}
\end{equation}
where $\phi$ is the directivity angle of antenna.  When $\phi = 2\pi$, it is an omnidirectional antenna with
$G = 1$.

When Alice transmits a message, the received signal power at Bob is given by
\begin{equation}\label{qq}
    P_{Rx} = A d_{a,b}^{-2}\exp (-Kd_{a,b})
\end{equation}
where $K$ is the overall absorption coefficient of the medium, $d_{a,b}$ is the distance between
Alice and Bob, and $A$ can be described as
\begin{equation}\label{22qq}
    A = P_{Tx}G_{Tx}G_{Rx}\frac{c^2}{16\pi^2f^2}=HG_{Tx}G_{Rx}
\end{equation}
where $P_{Tx}$ is the transmit power of Tx, $G_{Tx}$ and $G_{Rx}$ are the antenna gain of Tx and Rx,
respectively, $c$ is the speed of the EM wave, and $f$ is the operating frequency, $H =
P_{Tx}c^2/(16\pi^2f^2)$.

In addition to the path loss, any receiver will experience the Johnson-Nyquist noise generated by thermal
agitation of electrons in conductors \footnote {In this paper, we do not take into account the effect of the
molecular noise.}, which can be represented as follows
\begin{equation}\label{www}
    S_{JN}(f) = \frac{hf}{\exp(hf/k_BT)-1}
\end{equation}
where $h$ is Planck's constant, $k_B$ is Boltzmann constant, and $T$ is the temperature in Kelvin.

\subsubsection{Network and Blocking Model}
All transmitters form a stationary PPP $\Pi=\{X_i\}$ with the density $\lambda$ on the plane. Tx and Rx are
connected with LOS configurations, as is standard for a highly directional millimetre-wave or THz Band wireless
link through the air.

In a THz Band network, Rx suffers not only from the noise, but also from the aggregate interference from
other transmitters. However, due to the directionality of THz Band channels, nodes themselves may act as
blockers for interference. In this paper we use the blocking model proposed in \cite{SINR} to analyze the
aggregate interference. Suppose the interference of a certain interferer $J$ at the receiver Bob is zero if the
LOS path between $J$ and Bob is blocked by another interferer. For any interferer located at a distance $x$
from Bob, the blocking probability of the interference from this interferer can be estimated as follows
\begin{equation}\label{122}
    \mathbb{P}_B(x) = 1-\exp[-\lambda(x-r_B)r_B]
\end{equation}
where $r_B$ is the blocker radius, $\lambda$ is the network density.

Besides, if Bob is not in the coverage of the interferer $J$, then $J$ does not contribute to the aggregate
interference at Bob. Given Bob's antenna directivity angle $\phi$, the probability that the receiver is not in
coverage of an interferer can be given by
\begin{equation}\label{3424}
    \mathbb{P}_C = \frac{\phi}{2\pi}
\end{equation}
then the aggregate interference Bob experiences in a THz Band network can be represented as
\begin{equation}\label{ITHz}
    I^{(b)}_{THz} = A\sum^{\infty}_{i=1}r_i^{-2}\exp(-Kr_i)\cdot \mathbf{1}_{\{I_i>0\}}
\end{equation}
where $r_i$ is the distance between i-th interferer and Bob. $\mathbf{1}_{\{I_i>0\}}$ is an indicator function,
$\mathbf{1}_{\{I_i>0\}}=1$ if Bob is interfered by the interferer $i$, $\mathbf{1}_{\{I_i>0\}}=0$ if the
interference signal from $i$ is blocked, or Bob's antenna directivity is not in coverage of $i$. The probability
$\mathbb{P}\{\mathbf{1}_{\{I_i>0\}}=1\}=\mathbb{P}_C(1-\mathbb{P}_B)$.

\subsubsection{Rough Surface Scattering Model}
\begin{figure}
\centering \epsfig{file=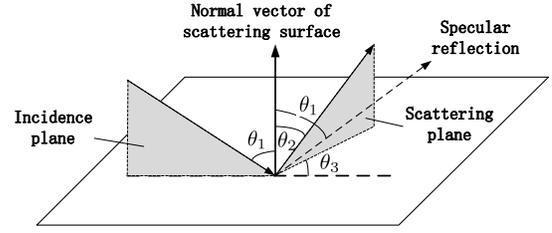, height=1.2in}
\caption{The model of scattering at a rough surface.}\label{Scattering}
\end{figure}

The general surface scattering model is
shown in Fig. \ref{Scattering}. A wave, which is incident on a rough surface under an angle $\theta_1$, is
scattered into the direction given by the angles $\theta_2$ and $\theta_3$.

\begin{figure}
\centering \epsfig{file=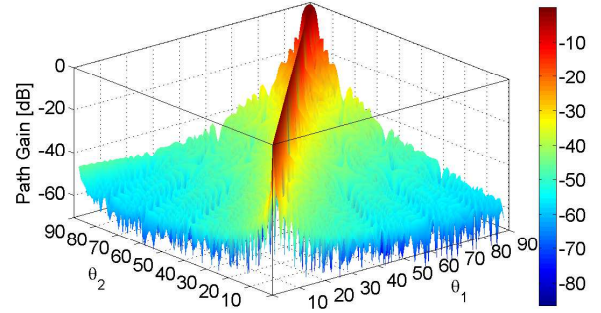, height=1.7in}
\caption{Path gain at 500 GHz frequency as a function of angles $\theta_1$ and $\theta_2$.
$\theta_1, \theta_2=0^{\circ}\cdots 90^{\circ}$ in steps of $1^{\circ}$, and $\theta_3$ is set to $0^{\circ}$.
The illuminated area is approximately 4cm$^2$, the surface correlation length $l_c=0.1$mm, the surface height variation $\sigma_h=0.01$mm.}\label{Scatteringss}
\end{figure}

For a rough surface with infinite conductivity, Kirchhoff scattering model \cite{Scattering_model} gives the
expression of the scattering path gain or scattering coefficient, $G(f, \sigma_h, l_c, \theta_1, \theta_2,
\theta_3)$, describing the scattered with respect to the incident power. In the expression of the Kirchhoff
approximation, parameters $l_c$ (the surface correlation length) and $\sigma_h$ (the standard deviation of the
surface height variation) describe the surface properties. For a smooth surface, the specular reflection
component always dominates. However a rough surface has a strong diffuse scattering contribution. Fig.
\ref{Scatteringss} shows the path gain at $f=500$GHz frequency as a function of angles $\theta_1$ and
$\theta_2$.

Kirchhoff scattering model offers a higher computational efficiency and can easily be implemented, so it is used
in many rough surface scattering papers \cite{Diffuse_1}\cite{Diffuse_2}\cite{Diffuse_3}, and we choose this
model to calculate the average scattered power.

\subsubsection{Assessment Metric}
We assume Willie place a receiver within the narrow beam from Alice to Bob. This set-up affords Willie
considerable high receiving signal strength, and the signal that he measures is large enough for him to detect the
communication between Alice and Bob. However, if Alice and Bob communication via NLOS configuration, what
Willie can obtain is a weak diffuse scattering field. To quantify the detection ability of Willie, we assess a
normalized secrecy capacity \cite{Nature}, which relates the strength of Willie's signal to Bob's signal as follows
\begin{equation}\label{cs}
    \bar{c}_s = \frac{\log(1+SINR_B)-\log(SINR_W)}{\log(1+SINR_B)}
\end{equation}
where $SINR_B$ and $SINR_W$ represent Bob and Willie's  signal to interference plus noise ratio on linear
scale, respectively. Given the reflecting path gain of Bob $G_B$ and the scattering path gain of Willie $G_W$,
$SINR_B$ and $SINR_W$ can be estimated as follows
\begin{eqnarray}
  SINR_B &=& \frac{A d_{a,b}^{-2}\exp (-Kd_{a,b})\cdot G_B}{S_{JN}(f)+I^{(b)}_{THz}} \\
  SINR_W &=& \frac{A d_{a,w}^{-2}\exp (-Kd_{a,w})\cdot G_W}{S_{JN}(f)+I^{(w)}_{THz}}
\end{eqnarray}
here $d_{a,b}$ is the length of NLOS reflecting path between Alice and Bob,  and $d_{a,w}$  the length of the
scattering path between Alice and Willie. $I^{(b)}_{THz}$ and $I^{(w)}_{THz}$ are the aggregate interferences
Bob and Willie experience, respectively.

The normalized secrecy capacity $\bar{c}_s$ is 1 if Willie receives no signal from Alice and 0 if Willie and Bob
receive the same signal. This quantity is a metric which is always used to define the security of a channel rather
than the covertness of wireless channels. However, we can use it to assess the likelihood of a successful covert
communication and design covert communication schemes to maximize it. If the normalized secrecy capacity
$\bar{c}_s$ is above a predefined threshold, we presume that covert communication is feasible.


\section{Covert Communication in AWGN Channels}\label{ch_3}
%


To transmit messages to Bob covertly and reliably, Alice should encode her messages. In this paper, we use the
classical encoder scheme used in \cite{square_law} and suppose that Alice and Bob have a shared secret of
sufficient length. At first, Alice and Bob leverage the shared secret and random coding arguments to generate a
secret codebook. Then Alice's channel encoder takes as input message of length $L$ bits and encodes them into
codewords of length $n$ at the rate of $R = L/n$ bits/symbol.

\subsection{Covertness}


In a dense wireless network, Willie not only experiences the background noise, but also the aggregate
interference from other transmitters. The total power of noise and interference Willie experiences can be
expressed as
\begin{equation}\label{eq_3}
   \sigma_w^2=\sigma^2_{w,0}+\sigma_{I_w}^2
\end{equation}
where $\sigma^2_{w,0}$ is the power of the background noise, $\sigma_{I_w}^2$ is the power of the
aggregate interference from other transmitters (defined in Eq. (\ref{eq_4_4})). In general, the
interference is more difficult to be predicted than the background noise, because the randomness of
aggregate interference comes from the randomness of PPP $\Pi$ and the fading channels, especially
in a mobile wireless network.

Let $\mathbb{P}_0$ be the joint probability density function (PDF) of $\mathbf{y}=\{y_i^{(w)}\}_{i=1}^n$ when
$\mathbf{H_0}$ is true, $\mathbb{P}_1$ be the joint PDF of $\mathbf{y}$ when $\mathbf{H_1}$ is true.
Using the same analysis method and the results from \cite{square_law}\cite{07096},
if Willie employs the optimal hypothesis test to minimize his probability of detection error
$\mathbb{P}_e^{(w)}$, then
\begin{equation}\label{eq_3_1}
   \mathbb{P}_e^{(w)}\geq\frac{1}{2}-\sqrt{\frac{1}{8}D(\mathbb{P}_1||\mathbb{P}_0)}
\end{equation}
where $D(\mathbb{P}_1||\mathbb{P}_0)$ is the relative entropy between $\mathbb{P}_1$ and
$\mathbb{P}_0$, and the lower bound of $\mathbb{P}_e^{(w)}$ can be estimated as follows \cite{07096}
\begin{eqnarray}\label{eq_4}
  \mathbb{P}_e^{(w)} &\geq & \frac{1}{2}-\sqrt{\frac{n}{8}}\cdot\frac{P_t\Psi_{a,w}}{2\sigma_w^2d^{\alpha}_{a,w}} \nonumber\\
  &=&  \frac{1}{2}-\sqrt{\frac{n}{8}}\cdot\frac{P_t\Psi_{a,w}}{2d^{\alpha}_{a,w}}\cdot\frac{1}{\sigma^2_{w,0}+\sigma_{I_w}^2} \nonumber\\
      &\geq & \frac{1}{2}-\sqrt{\frac{n}{8}}\cdot\frac{P_t\Psi_{a,w}}{2d^{\alpha}_{a,w}}\cdot\frac{1}{\sigma_{I_w}^2}
\end{eqnarray}
The last step is due to $\sigma^2_{w,0}\ll\sigma_{I_w}^2$, since in a dense and large-scale wireless network,
the background noise is negligible compared to the aggregate interference from other transmitters \cite{limited}.
Then the mean of $\mathbb{P}_e^{(w)}$ is
\begin{eqnarray}\label{eq_8}
    \mathbf{E}[\mathbb{P}_e^{(w)}]&\geq& \frac{1}{2}-\sqrt{\frac{n}{8}}\cdot\frac{P_t\mathbf{E}[\Psi_{a,w}]}{2d^{\alpha}_{a,w}}\cdot\mathbf{E}\biggl[\frac{1}{\sigma_{I_w}^2}\biggr] \nonumber\\
    &=& \frac{1}{2}-\sqrt{\frac{n}{8}}\cdot\frac{P_t}{2d^{\alpha}_{a,w}}\cdot\mathbf{E}\biggl[\frac{1}{\sigma_{I_w}^2}\biggr]
\end{eqnarray}
for all links experience unit mean Rayleigh fading.

To estimate $\mathbf{E}[1/\sigma_{I_w}^2]$, we should have the closed-form expression of the distribution of
$\sigma^2_{I_w}=\sum_{k\in\Pi}\frac{P_t}{d_{w,k}^{\alpha}}\Psi_{w,k}$. However, $\sigma^2_{I_w}$ is an
RV whose randomness originates from the random positions in PPP $\Pi$ and the fading channels. It obeys a
stable distribution without closed-form expression for its PDF or cumulative distribution function (CDF).  To
obtain the approximation of the mean of $1/\sigma_{I_w}^2$, we propose to use the Taylor expansion technique
(as discussed in \cite{SINR}, Appendix. B). Particularly, for the mean of an RV $Y=g(X)$, where $X$ is another
RV with mean $\mathbf{E}[X]$ and variance $\mathbf{Var}[X]$, we have
\begin{equation}\label{taylor}
    \mathbf{E}[Y] = g(\mathbf{E}[X]) + \frac{g''(\mathbf{E}[X])}{2}\cdot \mathbf{Var}[X]
\end{equation}

Next we estimate the mean and variance of $\sigma^2_{I_w}$. However, its mean is not exist if we employ the
unbounded path loss law (this may be partly due to the singularity of the path loss law at the origin). We then use
a modified path loss law to estimate the mean of $\sigma^2_{I_w}$,
\begin{equation}\label{eq_law}
    l(r)\equiv r^{-\alpha}\mathbf{1}_{r\geq\rho},~~r\in\mathbb{R}_+, ~~\text{for}~\rho\geq 0
\end{equation}

This law truncates around the origin and thus removes the singularity of impulse response function
$l(r)\equiv r^{-\alpha}$. The guard zone around the receiver (a ball of radius $\rho$) can be
interpreted as assuming any two nodes can't get too close. Strictly speaking, transmitters no
longer form a PPP under this bounded path loss law, but a hard-core point process in this case. For
relatively small guard zones, this model yields rather accurate results. For $\rho>0$, the mean and
variance of $\sigma^2_{I_w}$ are finite and can be given as \cite{Interference_Haenggi}
\begin{eqnarray}
  \mathbf{E}[\sigma^2_{I_w}] &=& \frac{\lambda d c_d}{\alpha-d}\mathbf{E}[\Psi]\mathbf{E}[P_t]\rho^{d-\alpha} \label{E}\\
  \mathbf{Var}[\sigma^2_{I_w}] &=& \frac{\lambda d c_d}{2\alpha-d}\mathbf{E}[\Psi^2]\mathbf{E}[P_t^2]\rho^{d-2\alpha} \label{var}
\end{eqnarray}
where $d$ is the spatial dimension of the network, the relevant values of $c_d$ are: $c_1=2$, $c_2=\pi$,
$c_3=4\pi/3$. In the following discussion, we assume $\rho = 1$, and all links experience unit mean Rayleigh
fading $\Psi\sim \mathrm{Exp}(1)$ with $\mathbf{E}[\Psi] = 1$ and $\mathbf{E}[\Psi^2] = 2$.

Therefore using Eq. (\ref{taylor}), (\ref{E}), and (\ref{var}), given the constant transmit power $P_t$,  $\mathbf{E}[1/\sigma_{I_w}^2]$ can be estimated as follows
\begin{eqnarray}
  \mathbf{E}\biggl[\frac{1}{\sigma_{I_w}^2}\biggr] &=&  \frac{1}{\mathbf{E}[\sigma^2_{I_w}]} + \frac{1}{(\mathbf{E}[\sigma^2_{I_w}])^3}\cdot \mathbf{Var}[\sigma^2_{I_w}] \nonumber \\
   &=&  \frac{\alpha-d}{\lambda d c_d P_t} + \biggl(\frac{\alpha-d}{\lambda d c_d P_t}\biggl)^3\cdot \frac{2\lambda d c_d}{2\alpha-d}\cdot P^2_t    \nonumber \\
   &=& \frac{\alpha-d}{\lambda d c_d P_t}\biggl[ 1+ \frac{2(\alpha-d)^2}{(2\alpha -d)dc_d}\cdot\frac{1}{\lambda}  \biggr ]   \nonumber \\
   &=& \frac{1}{P_t}\cdot f(\lambda) \label{26}
\end{eqnarray}
where $f(\lambda)$ is a function of $\lambda$,
\begin{equation}\label{Phi}
    f(\lambda)=\frac{1}{\lambda}\cdot\frac{\alpha-d}{d c_d }\biggl[ 1+ \frac{2(\alpha-d)^2}{(2\alpha -d)dc_d}\cdot\frac{1}{\lambda}  \biggr ]
\end{equation}

Thus, (\ref{eq_8}) and (\ref{26}) yield the lower bound of $\mathbf{E}[\mathbb{P}_e^{(w)}]$ as
\begin{eqnarray}\label{eq_15}
    \mathbf{E}[\mathbb{P}_e^{(w)}] &\geq & \frac{1}{2}-\sqrt{\frac{n}{8}}\cdot\frac{P_t}{2d^{\alpha}_{a,w}}\cdot\mathbf{E}\biggl[\frac{1}{\sigma_{I_w}^2}\biggr] \nonumber\\
         & = & \frac{1}{2}-\sqrt{\frac{n}{8}}\cdot\frac{P_t}{2d^{\alpha}_{a,w}}\cdot\frac{1}{P_t}\cdot f(\lambda) \nonumber\\
        & = & \frac{1}{2}-\sqrt{\frac{n}{8}}\cdot\frac{f(\lambda)}{2d^{\alpha}_{a,w}}
\end{eqnarray}

Suppose $\mathbf{E}[\mathbb{P}_e^{(w)}]\geq \frac{1}{2}-\epsilon$ for any $\epsilon>0$, then we
should set
\begin{equation}
\sqrt{\frac{n}{8}}\cdot\frac{f(\lambda)}{2d^{\alpha}_{a,w}}<\epsilon£¬
\end{equation}
and we have
\begin{equation}
d_{a,w}>\biggl[ \frac{1}{4\sqrt{2}\epsilon}\cdot f(\lambda) \biggr]^{1/\alpha}\cdot n^{1/(2\alpha)}.
\end{equation}

Therefore, as long as $d_{a,w}=\omega(n^{1/(2\alpha)})$, we can get $\mathbf{E}[\mathbb{P}_e^{(w)}]\geq
\frac{1}{2}-\epsilon$ for any $\epsilon>0$. This implies that there is no limitation on the transmit power $P_t$,
the critical factor is the distance between Alice and Willie.  This result is different from the works of Bash
\cite{square_law} and Soltani \cite{07096}, in which Alice's symbol power is a decreasing function of the
codeword length $n$. While this may appear counter-intuitive, the result in fact is explicable. We believe the
reasons are two folds. First, higher transmission signal power will create larger interference which will make
Willie more difficult to judge. Secondly, more close to the transmitter will give Willie more accurate estimation.
This theoretical result is also verified using the experimental results in Section \ref{ch_44}. Besides, the lower
bound of $d_{a,w}$ is related with the network density $\lambda$. If the network is sparse, the smaller the
density $\lambda$, the larger the function $f(\lambda)$, then the lower bound of $d_{a,w}$ will become
greater, which means that in the network settings with less interference, we should put Alice further away from
Willie to guarantee the covertness. In a denser network with a larger density, the lower bound of $d_{a,w}$
becomes smaller.

\subsection{Reliability}
Next, we estimate Bob's decoding error probability, denoted by $\mathbb{P}^{(b)}_e$. Let the total noise power
that Bob experiences be
\begin{equation}
\sigma^2_b=\sigma^2_{b,0}+\sigma^2_{I_b}
\end{equation}
where $\sigma^2_{b,0}$ is the power of background noise Bob observes, $\sigma_{I_b}^2$ is the power
of the aggregate interference from other transmitters. By utilizing the same
approach in \cite{square_law}, Bob's decoding error probability can be lower bounded as follows,
\begin{eqnarray}
  \mathbb{P}^{(b)}_e(\sigma^2_b) &\leq& 2^{nR-\tfrac{n}{2}\log_2\bigl(1+\tfrac{P_t}{2\sigma^2_b}\bigr)} \nonumber\\
  &=& 2^{nR-\tfrac{n}{2}\log_2\bigl[1+\tfrac{P_t}{2(\sigma^2_{b,0}+\sigma^2_{I_b})}\bigr]} \nonumber\\
   &=& 2^{nR}\biggl[1+\frac{P_t}{2(\sigma^2_{b,0}+\sigma^2_{I_b})}\biggr]^{-n/2} \nonumber\\
   &\leq & 2^{nR}\biggl[1+\frac{P_t}{2(\sigma^2_{b,0}+\sigma^2_{I_b})}\frac{n}{2}\biggr]^{-1}
\end{eqnarray}
where $R$ (bits/symbol) is the rate of encoder, and the last step is obtained by the following inequality
\cite{07096}
\begin{equation}
(1+x)^{-r}\leq (1+rx)^{-1} ,~~\text{for any}~~ r\geq 1~~ \text{and}~~ x>-1
\end{equation}

To estimate $\mathbb{P}^{(b)}_e(\sigma^2_b)$, we should have the closed-form expression of the distribution
of $\sigma^2_{I_b}$. However, $\sigma^2_{I_b}$ obeys a stable distribution without closed-form expression
for its PDF and CDF. To address wireless network capacity, Weber \emph{et al.} \cite{Weber} employed tools
from stochastic geometry to obtain asymptotically tight bounds on the distribution of the signal-to-interference
(SIR) level in a wireless network, yielding tight bounds on its complementary cumulative distribution function
(CCDF).

Define a random variable
\begin{equation}
\mathbf{Y} = \frac{\Sigma_{k\in\Pi}P_t\Psi_{k,b}d^{-\alpha}_{k,b}}{P_t\Psi_{a,b}d^{-\alpha}_{a,b}}=\frac{\sigma^2_{I_b}}{P_t\Psi_{a,b}d^{-\alpha}_{a,b}}
\end{equation}
then, the upper bound on the CCDF of RV $\mathbf{Y}$, $\bar{F}^u_\mathbf{Y}(y)$, can be expressed
as (\cite{Weber}, Eq. 27),
\begin{equation}
\bar{F}_\mathbf{Y}^u(y)=\frac{2}{2-\delta}\kappa\lambda y^{-\delta} + \mathcal{O}(y^{-2\delta})
\end{equation}
where $\kappa = \pi\mathbf{E}[\Psi^\delta]\mathbf{E}[\Psi^{-\delta}]\mathbf{E}[d^2_{a,b}]$, $\lambda$ is
the intensity of attempted transmissions in PPP $\Pi$, and $\delta=2/\alpha$. When $\Psi \sim
\mathrm{Exp}(1)$, $\kappa =
\pi\Gamma(1+\delta)\Gamma(1-\delta)d^2_{a,b}=\frac{\pi^2\delta}{\sin(\pi\delta)}d^2_{a,b}$.

Because ${\sigma^2_{I_b}}$ is a linear function of $\mathbf{Y}$ and they are positive correlation, we can get
the upper bound on CCDF of RV $\sigma^2_{I_b}$ as follows
\begin{eqnarray}\label{eq_5}
  \bar{F}^u_{{\sigma}^2_{I_b}}(x) &=& \mathbb{P}\{\sigma^2_{I_b} > x\} = \mathbb{P}\{P_t\Psi_{a,b}d_{a,b}^{-\alpha}\mathbf{Y}>x\} \nonumber\\
   &=& \mathbb{P}\{\mathbf{Y}>\frac{x}{P_t\Psi_{a,b}d_{a,b}^{-\alpha}}\} \nonumber\\
   &=& \frac{2}{2-\delta}\kappa\lambda\beta^{\delta} x^{-\delta} + \mathcal{O}(x^{-2\delta}) \nonumber\\
   &=& \eta\lambda\beta^{\delta} x^{-\delta} + \mathcal{O}(x^{-2\delta})
\end{eqnarray}
where $\eta =\frac{2}{2-\delta}\kappa$, $\beta=P_t\Psi_{a,b}d_{a,b}^{-\alpha}$. To strengthen the achievability results, we assume
the channel gain of channel between Alice and Bob is static and constant, $\mathbf{h}_{a,b}=1$.
Then $\beta$ can be denoted as $\beta=P_t d_{a,b}^{-\alpha}$.

Now define an RV $\bar{\sigma}^2_{I_b}$ who obeys the distribution of Eq. (\ref{eq_5}). Then we have
\begin{equation}\label{order}
    \mathbb{P}\{ \bar{\sigma}^2_{I_b} > x  \} > \mathbb{P}\{ \sigma^2_{I_b} > x  \}
\end{equation}
which implies that the RV $\bar{\sigma}^2_{I_b}$ stochastically dominates RV $\sigma^2_{I_b}$. According to the theory of stochastic orders \cite{Orders}\cite{Ordering},
\begin{equation}\label{mean}
   \mathbf{E}[g(\bar{\sigma}^2_{I_b})] > \mathbf{E}[g(\sigma^2_{I_b})]
\end{equation}
if $g(x)$ is non-decreasing.


Hence the upper bound of Bob's average decoding error probability can be estimated as follows
\begin{eqnarray}\label{eq_21}
  \mathbf{E}[\mathbb{P}^{(b)}_e(\sigma^2_b)] &\leq& \mathbf{E}\biggl[2^{nR}\biggl(1+\frac{nP_t/4}{\sigma^2_{b,0}+\sigma^2_{I_b}}\biggr)^{-1}\biggr] \nonumber\\
  & \overset{(a)}{<} & \mathbf{E}\biggl[2^{nR}\biggl(1+\frac{nP_t/4}{\sigma^2_{b,0}+\bar{\sigma}^2_{I_b}}\biggr)^{-1}\biggr] \nonumber\\
   &\overset{(b)}{=} & \int_0^{\infty}2^{nR}\biggl(1+\frac{nP_t/4}{\sigma^2_{b,0}+x}\biggr)^{-1}f^u_{\bar{\sigma}^2_{I_b}}(x)\mathrm{d}x \nonumber\\
   &=& 2^{nR}\int_{(\eta\lambda)^{\tfrac{1}{\delta}}\beta}^{\infty}\biggl(1+\frac{nP_t/4}{\sigma^2_{b,0}+x}\biggr)^{-1} \nonumber\\
     & & \times\eta\lambda\beta^\delta\delta x^{-(\delta+1)} \mathrm{d}x
\end{eqnarray}
here the inequality (a) holds because RV $\bar{\sigma}^2_{I_b}$ stochastically dominates RV
$\sigma^2_{I_b}$, and the function $g(x) = \bigl(1+\frac{nP_t/4}{\sigma^2_{b,0}+x}\bigr)^{-1}$ is
non-decreasing. In equation (b), $f^u_{\bar{\sigma}^2_{I_b}}(x)$ is PDF of RV $\bar{\sigma}^2_{I_b}$ whose
CCDF is expressed in Eq. (\ref{eq_5}). It's PDF can be represented as follows
\begin{equation}\label{eq_CDF}
f^u_{\bar{\sigma}^2_{I_b}}(x)=\eta\lambda\beta^\delta\delta x^{-(\delta+1)}, ~~x\in [(\eta\lambda)^{1/\delta}\beta, +\infty)
\end{equation}
where $\beta=P_t\Psi_{a,b}d_{a,b}^{-\alpha}$. We set $x\in [(\eta\lambda)^{1/\delta}\beta, +\infty)$ to normalize the function
so that it describes a probability density.

Let $a = nP_t/4$, the Eq. (\ref{eq_21}) can be calculated as follows
\begin{eqnarray}\label{eq_22}
 & &  \mathbf{E}[\mathbb{P}^{(b)}_e(\sigma^2_b)]  \nonumber \\
  &<& 2^{nR}\int_{(\eta\lambda)^{\tfrac{1}{\delta}}\beta}^{\infty}\biggl(1+\frac{a}{\sigma^2_{b,0}+x}\biggr)^{-1}\eta\lambda\beta^\delta\delta x^{-(\delta+1)} \mathrm{d}x  \nonumber\\
    &=& 2^{nR}\eta\lambda\beta^\delta\delta\biggl[\frac{\pi a}{(a+\sigma^2_{b,0})^{3/2}}- \frac{2a\tan^{-1}\bigl(\tfrac{\eta\lambda\beta^\delta}{\sqrt{a+\sigma^2_{b,0}}}\bigr)}{(a+\sigma^2_{b,0})^{3/2}} \nonumber\\
    & & +\frac{2\sigma^2_{b,0}}{\eta\lambda\beta^\delta(a+\sigma^2_{b,0})}\biggr]. \nonumber
\end{eqnarray}

When $n$ is large enough, we have
\begin{equation}
  a = nP_t/4\gg \sigma^2_{b,0}, ~~ a+\sigma^2_{b,0}\approx a.
\end{equation}

Let the path loss exponent $\alpha=4$, $\delta=1/2$, $\eta =\frac{2}{2-\delta}\kappa
=\frac{2}{2-\delta}\frac{\pi^2\delta}{\sin(\pi\delta)}d_{a,b}^2=\frac{2\pi^2}{3}d_{a,b}^2$, we have
\begin{equation}\label{asasa}
    \frac{2a\tan^{-1}\bigl(\tfrac{\eta\lambda\beta^\delta}{\sqrt{a+\sigma^2_{b,0}}}\bigr)}{(a+\sigma^2_{b,0})^{3/2}}  > \frac{2\sigma^2_{b,0}}{\eta\lambda\beta^\delta(a+\sigma^2_{b,0})}
\end{equation}
provided that the transmit power $P_t$ satisfies the following condition\footnote{This inequality can be easily
derived, mainly because $\lim_{n\rightarrow\infty}\sqrt{n}\tan^{-1}(\frac{c}{\sqrt{n}}) = c$ for a given
constant $c$, and $a+\sigma^2_{b,0}\approx a$ when $n\rightarrow\infty$.},
\begin{equation}\label{wr}
    P_t > \frac{9}{4\pi^4\lambda^2\Psi_{a,b}}\cdot\sigma^2_{b,0}
\end{equation}

Therefore we have
\begin{eqnarray}
  \mathbf{E}[\mathbb{P}^{(b)}_e(\sigma^2_b)] & \overset{(a)}{<} & 2^{nR}\eta\lambda\beta^\delta\delta\biggl[\frac{\pi a}{(a+\sigma^2_{b,0})^{3/2}}\biggr] \nonumber\\
   &\overset{(b)}{<}& 2^{nR}\eta\lambda\beta^\delta\delta  \frac{\pi}{\sqrt{a}} \nonumber\\
   &<& 2^{nR}\eta\lambda\beta^\delta\delta\frac{2\pi}{\sqrt{nP_t}} \nonumber\\
   &=& 2^{nR}\frac{2\pi^2}{3}d_{a,b}^2\lambda P_t^{1/2}\mathbf{E}[\Psi^{1/2}]d_{a,b}^{-\alpha/2}\delta\frac{2\pi}{\sqrt{nP_t}} \nonumber\\
   &=& 2^{nR}\frac{\pi^{7/2}\lambda}{3\sqrt{n}}
\end{eqnarray}
where the inequality (a) holds because we have Eq. (\ref{asasa}), (b) is due to $a+\sigma^2_{b,0}\approx a$.
$\mathbf{E}[\Psi^{1/2}]=\Gamma(1+1/2)=\sqrt{\pi}/2$ for $\Psi\sim \mathrm{Exp}(1)$.

Let $\mathbf{E}[\mathbb{P}^{(b)}_e(\sigma^2_b)]\leq\epsilon$ for any $\epsilon>0$, we have
\begin{equation}
  nR\leq \log_2\biggl(\frac{3\epsilon}{\pi^{7/2}\lambda}\cdot\sqrt{n}\biggr)
\end{equation}
which implies that Bob can receive
\begin{equation}
L=\mathcal{O}(\log_2\sqrt{n})~~ \text{bits}
\end{equation}
reliably in $n$ channel uses in the case that $\alpha=4$, and $L$ decreases as the density of interferers
$\lambda$ become larger. This may be a pessimistic result at first glance since it is lower than the bound derived
by Bash \cite{square_law}, i.e., Bob can reliably receive $\mathcal{O}(\sqrt{n})$ bits in $n$ channel uses. This is
reasonable because Bob experiences not only the background noise but also the aggregate interference, resulting
lower transmit throughput. However, in the work of Bash, Alice's symbol power is a decreasing function of the
codeword length $n$, i.e., her average symbol power $P_f\leq\frac{cf(n)}{\sqrt{n}}$. When Bob use
threshold-scheduling scheme to receive signal, Bob will have higher outage probability as $n\rightarrow\infty$.
This is because Alice's symbol power will become very lower to ensure the covertness as $n\rightarrow\infty$.
If we hide communications in noisy wireless networks, the spatial throughput is higher than the work of Bash in
which only background noise is considered. This will be discussed in Section \ref{ch_44}.

\subsection{Discussions}\label{ch_44}
\subsubsection{Spatial Throughput} The spatial throughput is the expected spatial density of successful
transmissions in a wireless network \cite{Weber}
\begin{equation}
\tau(\lambda)=\lambda(1-q(\lambda))
\end{equation}
where $q(\lambda)$ denotes the probability of transmission outage when the intensity of attempted
transmissions is $\lambda$ for given SINR requirement $\xi$.

In the work of Bash \emph{et al.} \cite{square_law}, only background noise is taken into account, Alice can
transmit $\mathcal{O}(\sqrt{n})$ bits reliably and covertly to Bob over $n$ uses of AWGN channels. To achieve
the covertness, Alice must set her average symbol power $P\leq\frac{cf(n)}{\sqrt{n}}$. Soltani \emph{et
al.}\cite{jammer2}\cite{07096} further expanded Bash's work. They introduced the friendly node closest to
Willie to produce artificial noise. This method allows Alice to reliably and covertly send
$\mathcal{O}(\min\{n,\lambda^{\alpha/2}\sqrt{n}\})$ bits to Bob in $n$ channel uses when there is only one
adversary. In their network settings, Alice must set her average symbol power
$P_a=\mathcal{O}(\frac{c\lambda^{\alpha/2}}{\sqrt{n}})$ to avoid being detected by Willie. Thus, given an
SINR threshold $\xi$, $\sigma_{b,0}^2\geq 1$, and Rayleigh fading with $\Psi\sim\mathrm{Exp}(1)$, the
outage probability  of Soltani's method is
\begin{eqnarray}
  q^J(\lambda) &=& \mathbb{P}\biggl\{\mathrm{SINR}=\frac{P_a\Psi d^{-\alpha}_{a,b}}{\sigma^2_{b,0}+P_f\Psi d^{-\alpha}_{a,f}}<\xi\biggr\} \nonumber \\
    &\geq&  \mathbb{P}\{P_a\Psi d^{-\alpha}_{a,b}<\xi\} \nonumber \\
    &\geq& \mathbb{P}\biggl\{\frac{c\lambda^{\alpha/2}}{\sqrt{n}}\Psi d^{-\alpha}_{a,b}<\xi\biggr\} \nonumber \\
    &=& \mathbb{P}\biggl\{\Psi<\frac{1}{c\lambda^{\alpha/2}}d^{\alpha}_{a,b}\xi\sqrt{n}\biggr\} \nonumber \\
    &=& 1-\exp\biggl\{-\frac{1}{c\lambda^{\alpha/2}}d^{\alpha}_{a,b}\xi\sqrt{n}\biggr\}
\end{eqnarray}
where $P_f$ is the jamming power of the friendly node, and $d_{a,f}$ is the distance between Alice
and the friendly node. Then the spatial throughput of the network is
\begin{equation}\label{eq_spatial_J}
\tau^J(\lambda)=\lambda(1-q^J(\lambda))\leq\lambda\exp\biggl\{-\frac{1}{c\lambda^{\alpha/2}}d^{\alpha}_{a,b}\xi\sqrt{n}\biggr\}.
\end{equation}

If we hide communications in the aggregate interference of a noisy wireless network with
randomized transmissions in Rayleigh fading channel and the SINR threshold is set to $\xi$, the
spatial throughput is \cite{Weber}
\begin{equation}\label{eq_spatial_I}
\tau^I(\lambda)=\lambda\exp\{-\pi\lambda\xi^{\delta}d^2_{a,b}\Gamma(1+\delta)\Gamma(1-\delta)\}
\end{equation}
where $\delta=2/\alpha$.

As a result of Eq. (\ref{eq_spatial_J}) and (\ref{eq_spatial_I}), we can state that, by using a friendly jammer
near Willie to help Alice, Alice can reliably and covertly send
$\mathcal{O}(\min\{n,\lambda^{\alpha/2}\sqrt{n}\})$ bits to Bob in $n$ channel uses, which is higher than
$\mathcal{O}(\log_2\sqrt{n})$ bits when the aggregate interference is involved. But as $n\rightarrow\infty$,
the spatial throughput of the jamming scheme $\tau^J(\lambda)$ reduces to zero, and the covert
communication hiding in interference can achieve a constant spatial throughput $\tau^I(\lambda)$ which is higher
than $\tau^J(\lambda)$. Hence, although this approach has lower covert throughput for any pair of nodes, it has a
considerable higher throughput from the network perspective.


\subsubsection{Interference Uncertainty} From the analysis above, we found that the interference can indeed
increase the privacy throughput. If we can deliberately deploy interferers to further increase the interference
Willie experiences and does not harm the receiver, the security performance can be enhanced, such as the
methods discussed in \cite{jammer}\cite{jammer1}\cite{jammer2}\cite{INFOCOM_2016}.

\begin{figure}
\centering \epsfig{file=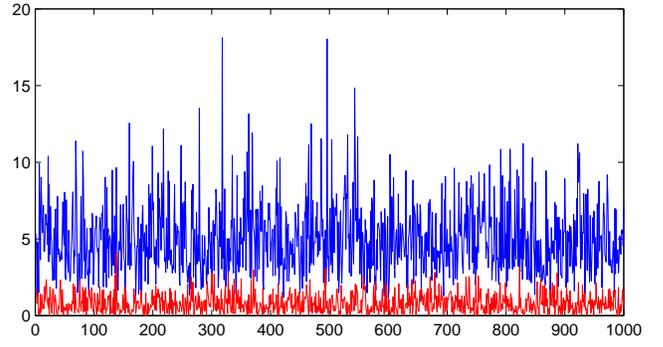, height=1.8in}
\caption{Sequences of 1000 realizations of noise and aggregate interference. Here a bounded path
loss law is used, $l(x)=\min\{1, r^{-\alpha}\}$. The transmit power $P_t$ of nodes are all unity,
links experience unit mean Rayleigh fading, $\Psi\sim\mathrm{Exp}(1)$, and $\alpha=4$. A reference point is located at the center of a square area 100m$\times$100m.
Interferers deployed in this area form a PPP on the plane with $\lambda=1$.
Interference the reference point sees is depicted in blue, the noise is depicted in red.}\label{Interference}
\end{figure}

Overall, the improvement comes from the increased interference uncertainty. If there is only noise
from Willie's surroundings, Willie may estimate the noise level by gathering samples although the
background noise can be unpredictable to some extent. However, the aggregate interference is more
difficult to be predicted. Fig.\ref{Interference} illustrates this situation by sequences of
realizations of the noise (normal distribution with the variance one) and the aggregate
interference. The interference has greater dispersion than the
background noise, thus it is more difficult to sample interferences to obtain a proper interference
level.

Additionally, the aggregate interference is always dominated by the interference generated by the nearest
interferer. If an interferer gets closer to Willie than Alice, Willie will be overwhelmed by the signal of the
interferer, and his decision will be uncertain. Let $r_1$ be the distance between the nearest interferer and
Willie, $f_{R_1}(r)$ be the PDF of the nearest-neighbor distance distribution on the plane $\mathbb{R}^2$
\cite{distances}, then
\begin{eqnarray}
  \mathbb{P}\{r_1<d_{a,w}\} &=& \int_0^{d_{a,w}}f_{R_1}(r)\mathrm{d}r \nonumber \\
   &=& \int_0^{d_{a,w}}2\pi\lambda r\exp(-\pi\lambda r^2)\mathrm{d}r \nonumber \\
   &=& 1-\exp(-\pi\lambda d_{a,w}^2).
\end{eqnarray}
We see that when $d_{a,w}=1$ and $\lambda=1$, $\mathbb{P}\{r_1<d_{a,w}\} = 0.9568$ - that is, there
is a dramatically high probability that Willie will experience more interference from the nearest
interferer. He will confront a dilemma to make a binary decision. In a dense and noisy wireless
network, Willie cannot determine which node is actually transmitting if he cannot get closer than
$\Theta(n^{1/(2\alpha)})$ and cannot sure no other nodes located in his detect region.

\subsubsection{Practical Method and Experimental Results} In the previous analysis, when Willie samples the
noise to determine the threshold of his detector (radiometer), we presuppose that Willie knows whether Alice is
transmitting or not, and he knows the power level of $\sigma^2_{I_w}$. In practice, Willie has no prior
knowledge on whether Alice is transmitting or not during his sampling process. This implies that Willie's sample
$y^{(w)}_i$ follows the distribution
\begin{equation}\label{eq_H111}
    y_i^{(w)} \sim \mathcal{N}\biggl(\sqrt{\frac{P_t}{d_{w,a}^{\alpha}}}\mathbf{h}\cdot s_i\cdot\mathbf{1}_A+\sum_{k\in\Pi}\sqrt{\frac{P_t}{d_{w,k}^{\alpha}}} \mathbf{h}\cdot s_i^{(k)},\sigma^2_{w,0}\biggr)
\end{equation}
where $\mathbf{1}_A$ is an indicator function, $\mathbf{1}_A=1$ if Alice is transmitting,
$\mathbf{1}_A=0$ if Alice is silent, and the transmission probability
$\mathbb{P}\{\mathbf{1}_A=1\}=p$.

If Alice transmits messages and is silent alternately, Willie cannot be certain whether the samples contain Alice's
signals or not. To confuse Willie, Alice should not generate burst traffic, but transforming the bulk message into
a smooth network traffic with transmission and silence alternatively. She can divide the time into slots, then put
message into small packets. After that, Alice sends a packet in a time slot with the transmission probability $p$
and keeps silence for the next slot with the probability $1-p$, and so on. Via this scheduling scheme, Alice can
guarantee that Willie's samples are the mix of noise and signal which are undistinguishable by Willie.

\begin{figure}
\centering \epsfig{file=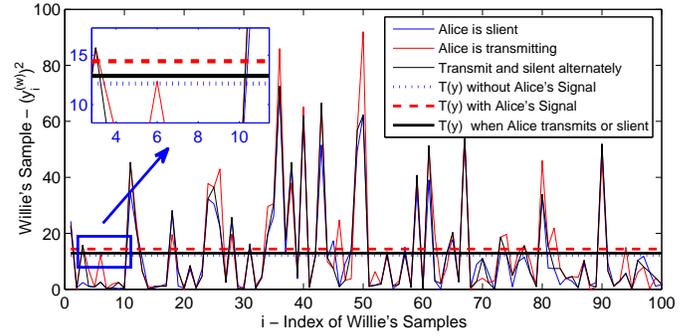, height=1.8in}
\caption{Sequences of 100 Willie's samples $[y_1^{(w)}]^2,..., [y_n^{(w)}]^2$ in the cases that Alice is silent, transmitting, or transmit and silent alternately. $T(\mathbf{y})=\frac{1}{n}\sum^n_{k=1}[y_k^{(w)}]^2$ in three cases are depicted as three lines. Here a bounded path loss law is used, $l(x)=\min\{1, r^{-\alpha}\}$. The transmit power $P_t$ is unity,
links experience unit mean Rayleigh fading, $\Psi\sim\mathrm{Exp}(1)$, $\alpha=4$, and $\sigma^2_{w,0}=1$. Willie is located at the center of a square area 100m$\times$100m. The distance between Alice and Willie $d_{a,w}=1$. Interferers deployed in this area form a PPP on the plane with $\lambda=1$.
}\label{Willie_samples}
\end{figure}

Next we provide an experimentally-supported analysis of this method. Fig. \ref{Willie_samples}
illustrates an example of sequences of 100 Willie's samples $[y_1^{(w)}]^2,..., [y_n^{(w)}]^2$ in
the case that Alice is silent, transmitting, or transmitting and silent alternately. Willie then
computes $T(\mathbf{y})=\frac{1}{n}\sum^n_{k=1}[y_k^{(w)}]^2$. Clearly, when Alice alternates
transmission with silence, Willie's sample value $T(\textbf{y})$ will decrease and quite near the
value when Alice is silent. The transmitted signals resemble white noise, and are sufficiently weak in this way.

\begin{figure}
\centering \epsfig{file=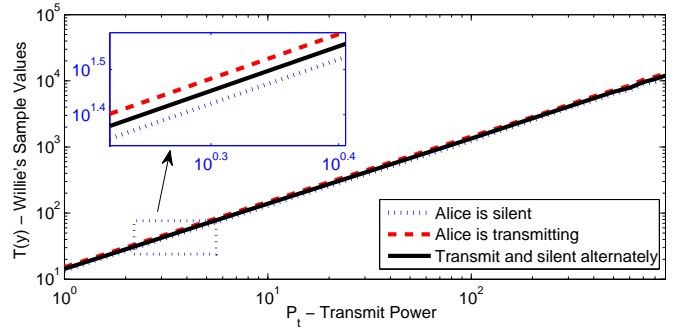, height=1.8in}
\caption{The transmit power $P_t$ versus Willie's sample values $T(\textbf{y})$ which are the average of 100 experiment runs, each with the number of samples $n=500$.
During each run of simulation, to obtain a sample $y_i^{(w)}$, a random wireless network obeying PPP on the plane is generated.
Here the distance between Alice and Willie $d_{a,w}=1$. All nodes in the network use constant transmit power.}\label{pt_y_mean}
\end{figure}

With the same simulation settings of Fig. \ref{Willie_samples}, we evaluate Willie's sample values
$T(\textbf{y})$ by varying the transmit power $P_t$. As displayed in Fig. \ref{pt_y_mean}, the
value $T(\textbf{y})$ changing with $P_t$ is displayed in three cases, i.e., Alice is transmitting,
silent, as well as transmitting and silent alternately. We find that when Alice employs the
alternation method, Willie's sample values decrease, approximating to the case Alice is silent.
Further, the results indicate that higher transmit power cannot lead to stronger capability for
Willie to distinguish Alice's transmission behavior. With the transmit power increases, Willie's
sample values $T(\textbf{y})$ increase. However, the aggregate interference increases as well,
resulting in the gap of sample values between Alice's transmission and silence does not increase.
Consequently, this is consistent with the result of the previous analysis, which indicates that increasing the
transmit power $P_t$ does not increase the risk of being detected by Willie in a wireless network.

\begin{figure}
\centering \epsfig{file=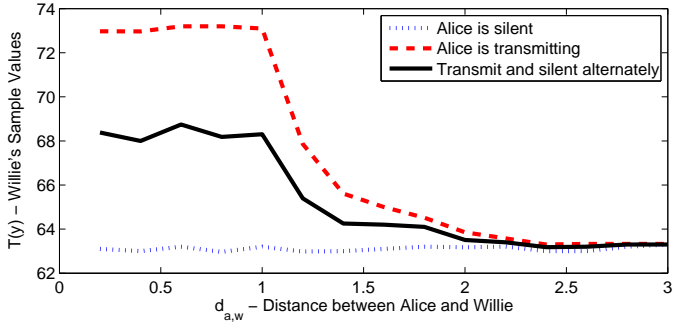, height=1.8in}
\caption{The distance between Alice and Willie $d_{a,w}$ versus Willie's sample values $T(\textbf{y})$ which are the average of 100 experiment runs,
each with the number of samples $n=500$. Here the transmit power $P_t=10$, and the transmission
probability $p=0.5$.}\label{daw_y_mean_1}
\end{figure}

Further, one of critical factors affecting covert communication is the
parameter $d_{a,w}$, the distance between Alice and Willie, which should satisfies
$d_{a,w}=\omega(n^{1/(2\alpha)})$ to ensure communication covertly. Fig. \ref{daw_y_mean_1}
illustrates Willie's sample values $T(\textbf{y})$ by varying the distance $d_{a,w}$. As the
results show, when Alice is silent, Willie's sample values $T(\textbf{y})$ barely change since
Willie only experiences the background noise and aggregate interference. When Alice is transmitting,
persistence or alteration, Willie's sample values increase with decreasing the distance $d_{a,w}$.
When $d_{a,w}\leq 1$, Willie's sample values become relatively stable since we employ the bounded
path loss law $l(x)=\min\{1, r^{-\alpha}\}$.

\begin{figure} \centering
\subfigure[n=1000]{ \epsfig{file=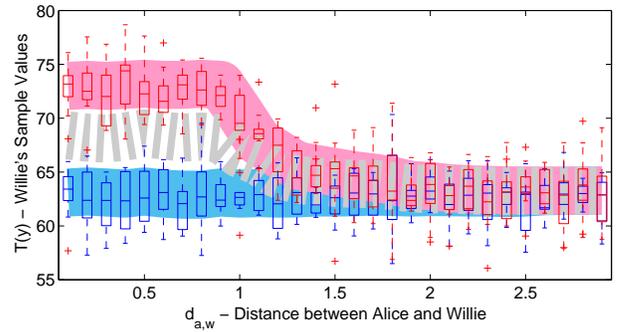, height=1.8in }}
\subfigure[n=3000]{ \epsfig{file=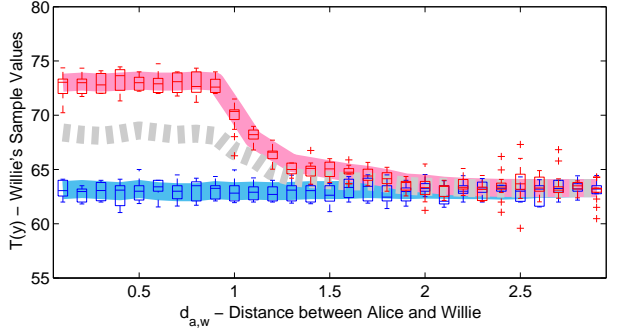, height=1.8in} }
\caption{The discreteness of Willie's sample values $T(\textbf{y})$ versus the distance $d_{a,w}$ when the number of samples $n=1000$ and $n=3000$.
At each subfigure, three simulation curves are given (from top to bottom): Alice is transmitting, transmitting and silent alternately (with transmission
probability $p=0.5$), and silent completely.
For each occasion, given a value $d_{a,w}$ and $n$, we implement 20 experiment runs to obtain 20 sample values $T(\textbf{y})$, and depict the discreteness of $T(\textbf{y})$ in boxplot form.
The width of curves also represent the dispersion degree of Willie's sample values.}
\label{boxplot}
\end{figure}

For the following analysis, we evaluate the effect of the number of samples $n$ on the distance
between Alice and Willie $d_{a,w}$. We start by comparing Willie's sample values by varying $n$ to
show the difference in performance. The results in Fig. \ref{boxplot} shows $T(\textbf{y})$ with
respect to the distance $d_{a,w}$ when $n=1000$ and $n=3000$. As can be seen, although the curves
of the average $T(\textbf{y})$ do not change, the discreteness of $T(\textbf{y})$ decreases with
increasing the number of samples $n$. As to Willie, to detect Alice's transmission attempts, he
should distinguish the three lines in the picture with relatively low probability of error. The
only way to decrease the probability of error is increasing the number of samples. By choosing a
larger value for $n$, Willie's uncertainty on noise and interference decreases, hence he can stay
far away from Alice to detect her transmission attempt. As illustrated in Fig. \ref{boxplot}(a),
Willie cannot distinguish Alice's transmission from silence when he stays at a distance of more
than 1 meter to Alice. However, when Willie increases the number of samples, he can distinguish
Alice's behavior far away. As depicted in Fig. \ref{boxplot}(b), Willie can detect Alice's
transmission at the distance between 1 and 1.5 meters with low probability of error. Overall, this
experimental result agrees with the previous theoretical derivation, i.e., given
the value $n$, the distance between Alice and Willie should be larger than a bound to ensure the
covertness, and the bound of $d_{a,w}$ increases with the increasing of $n$.

\section{Covert Communications in THz Band Communication Networks}
In THz Band, we always use directional communication channels with a beam divergence angle much smaller
than that used by existing mobile networks, which often use $120^{\circ}$ sectors. Intuitively, the narrow,
razor-sharp beam of THz Band can drastically limit the eavesdropping probability and can improve the data security \cite{AKYILDIZ201416}.
However, as discussed in \cite{Nature}, an eavesdropper can place an object in the path of the THz Band
transmission to scatter radiation towards the eavesdropper. Even when Alice and Bob are transmitted at high
frequencies with very directional and narrow beam, the eavesdropper can intercept signals in their LOS directional
transmissions.

Covert communication is more difficult than anti-eavesdropping communication. Next we present a scheme that
utilizes reflection or diffuse scattering from the rough surface to prevent being detected by Willie.

\subsection{Covert Communications in THz Band Networks}
In a THz Band network (as depicted in Fig. \ref{thz_model}), we suppose that all transmitters form a stationary
PPP $\Pi=\{X_i\}$ with the density $\lambda$ on the plane. Alice and Bob have a LOS highly directional mm-wave
or THz Band wireless link. Willie is located in the THz Band transmission path between Alice and Bob, and
evaluates the signal strength. Willie's goal is to detect the transmit behavior between Alice and Bob.


From Willie's point of view, his best policy is inserted himself into the LOS transmission path between Alice and
Bob. Therefore, to bypass the detection of Willie, Alice and Bob cannot use the LOS wireless link between them.
To perform covert communications, Alice and Bob need resorting to reflection or diffuse scattering NLOS
transmission paths,
\begin{itemize}
  \item \textbf{Specular Reflection Covert Communication}: At first, Alice and Bob find a surface in the
      surroundings that the THz beam from Alice can be specularly reflected to the antenna of Bob, i.e., the
      specular reflection path $\overrightarrow{AO_1}$ and $\overrightarrow{O_1B}$ in Fig. \ref{thz_model}.
      As illustrated in Fig. \ref{Scatteringss},  among the signals scattered from a rough surface,  the specular
      reflection component always dominates.
  \item \textbf{Diffuse Scattering Covert Communication}: If a specular reflection path cannot be found,
      Alice and Bob find a diffuse scattering path so that Bob's received signal strength is above a threshold,
      such as the diffuse scattering path $\overrightarrow{AO_2}$ and $\overrightarrow{O_2B}$ in Fig.
      \ref{thz_model}. Although the diffuse scattering signal is weak in comparison to the specular component, it
      is sometimes still high enough to enable the NLOS link on short distances.
\end{itemize}

\subsection{Analysis}
We use normalized secrecy capacity $\bar{c}_s$ to assess the likelihood of a successful covert communication.
If $\bar{c}_s$ is above a predefined threshold, we presume that the covert communication is feasible. To
estimate $\bar{c}_s$, we need to calculate the aggregate interference $I^{(b)}_{THz}$, $I^{(w)}_{THz}$ and
$SINR_B$, $SINR_W$. Next, we first estimate the mean of the aggregate interference $I^{(b)}_{THz}$ Bob
observed (in Eq. (\ref{ITHz})) as follows,
\begin{eqnarray}
  \mathbf{E}[I^{(b)}_{THz}]&=& \mathbf{E}\biggl[A\sum^{\infty}_{i=1}r_i^{-2}e^{-Kr_i}\cdot \mathbf{1}_{\{I_i>0\}}\biggr] \nonumber\\
   &\overset{(a)}{=}& A\lambda \int_{X_i\in \Pi}r_i^{-2}e^{-Kr_i}\cdot \mathbb{P}_C(1-\mathbb{P}_B)\mathrm{d}X_i \nonumber\\
   &=& A\lambda \int^{2\pi}_0\mathrm{d}\theta \int^R_{r_B}r^{-2}e^{-Kr}\frac{\phi}{2\pi}\cdot e^{-\lambda(r-r_B)r_B} r\mathrm{d}r\nonumber\\
   &=& A\lambda\phi e^{\lambda {r_B}^2}\int^R_{r_B}\frac{1}{r}e^{-(K+\lambda r_B)r} \mathrm{d}r\nonumber\\
   &=& A\lambda\phi e^{\lambda {r_B}^2}\biggl[Ei(-R(K+\lambda r_B)) \nonumber\\
   & & - Ei(-r_B(K+\lambda r_B))  \biggr]
\end{eqnarray}
where $Ei(\cdot)$ is the exponential integral function. $R$ is the radius of the zone where the nodes provide
non-negligible interferences. The signal that comes from Tx that is further than R is considered as the background noise. Eq. (a)
follows directly after Campbell's theorem \cite{Interference_Haenggi} for the mean of a sum function of a
stationary PPP $\Pi=\{X_i\}$.

Similarly, the variance of the aggregate interference $I^{(b)}_{THz}$ can be obtained as follows,
\begin{eqnarray}
  & & \mathbf{Var}[I^{(b)}_{THz}] \nonumber \\
  &=& \mathbf{Var}\biggl[\sum^{\infty}_{i=1}Ar_i^{-2}e^{-Kr_i}\cdot \mathbf{1}_{\{I_i>0\}}\biggr] \nonumber\\
   &\overset{(a)}{=}& \lambda \int_{X_i\in \Pi}\bigl[Ar_i^{-2}e^{-Kr_i}\cdot \mathbb{P}_C(1-\mathbb{P}_B)\bigr]^2 \mathrm{d}X_i \nonumber\\
   &=& \lambda \int^{2\pi}_0\mathrm{d}\theta \int^R_{r_B}\bigl[Ar^{-2}e^{-Kr}\cdot\frac{\phi}{2\pi}\cdot e^{-\lambda(r-r_B)r_B}\bigr]^2 r\mathrm{d}r\nonumber\\
   &=& A^2\lambda\frac{\phi^2}{2\pi} e^{2\lambda {r_B}^2}\int^R_{r_B}\frac{1}{r^3}e^{-(2K+2\lambda r_B)r} \mathrm{d}r\nonumber\\
   &=& A^2\lambda\frac{\phi^2}{2\pi} e^{2\lambda {r_B}^2}\biggl\{  2(K+\lambda r_B)^2Ei\bigl[-2(K+\lambda r_B)  \bigr] \nonumber\\
    & & + e^{-2(K+\lambda r_B)r}\cdot \bigl(\frac{K+\lambda r_B}{r}-\frac{1}{2r^2}\bigr)  \biggr\}\biggl\arrowvert^R_{r_B}
\end{eqnarray}
Here Eq. (a) also follows directly after Campbell's theorem for the variance of a sum function of a stationary
PPP $\Pi=\{X_i\}$.

With the mean and variance of $I^{(b)}_{THz}$ in hands, we can estimate the mean of $SINR_B$ by Taylor
expansion technique in \cite{SINR} as follows,
\begin{eqnarray}
  \mathbf{E}[SINR_B] &=& \mathbf{E}\biggl[ \frac{P_{Rx}}{S_{JN}(f)+I^{(b)}_{THz}} \biggr]  \\
   &=& \frac{P_{Rx}}{S_{JN}(f)+\mathbf{E}[I^{(b)}_{THz}]}  \nonumber\\
   & &  + \frac{P_{Rx}}{(S_{JN}(f)+\mathbf{E}[I^{(b)}_{THz}])^3}\cdot \mathbf{Var}[I^{(b)}_{THz}] \nonumber
\end{eqnarray}
Here $P_{Rx}$ is the received signal strength of Bob, $P_{Rx} = A d_{a,b}^{-2}\exp (-Kd_{a,b})\cdot G_B$,
$G_B$ is the reflecting or scattering path gain of Bob, which is obtained from Kirchhoff scattering model
\cite{Scattering_model}, and $S_{JN}(f)$ is the Johnson-Nyquist noise, described in Eq. (\ref{www}).  Similarly,
we can get the approximation of the mean of $SINR_W$.

Next we assess the effects of the operating frequency, network density, the surface roughnesses, and the
scattering angle of Willie on the mean of normalized secrecy capacity $\bar{c}_s$. Throughout this section, we
assume that interference coming from the nodes more than $R=10$m from the receiver is zero, the coefficient
$H$ introduced in Eq. (\ref{22qq}) is set to 1. The blocker radius of every node is $r_B = 0.1$m, the distance
between Alice and Bob is $d_{a,b} = 5$m, the absorption coefficient is assume to be a constant $K = 0.01$. All
nodes in the network are equipped with directional antennae (Tx and Rx) with directivity angle $\phi = \pi/18$.
Also, throughout this section we assume the illuminated area of the reflection surface is approximately 4cm$^2$,
the surface correlation length $l_c=1.8$mm.

\subsubsection{The Effect of Network Density $\lambda$}
\begin{figure}
\centering \epsfig{file=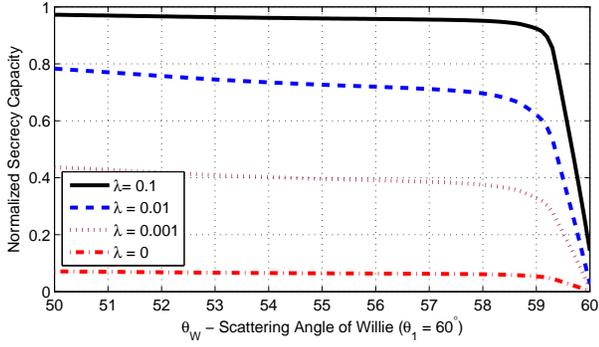, height=1.8in}
\caption{The normalized secrecy capacity $\bar{c}_s$ versus the scattering angle of Willie $\theta_W (50^{\circ}\cdots 60^{\circ})$ for different network density $\lambda$.
Here the incidence angle of Alice $\theta_1 = 60^{\circ}$, the surface height variation $\sigma_h=0.088$mm, and the operating frequency $f=500$GHz.}\label{lambda}
\end{figure}

At first we analyze the effect of network density $\lambda$ on the normalized secrecy capacity $\bar{c}_s$.
As illustrated in Fig. \ref{lambda}, if the incident angle of Alice $\theta_1 = 60^{\circ}$ and Bob's antenna is
located exactly in the specular reflection direction of Alice's signal, the closer Willie's scattering angle
$\theta_W$ to $\theta_1$, the smaller secrecy capacity $\bar{c}_s$ we can get. This is obvious because the
scattering coefficients $G_B$  and $G_W$ are very close when $\Delta = \theta_1 - \theta_W$ is small. On
the other hand, the higher the network density $\lambda$, the larger the normalized secrecy capacity and the
covert communications are more likely to succeed. Indeed, as illustrated in Fig. \ref{lambda}, if there is no other
interferers in the surroundings ($\lambda = 0$), the normalized secrecy capacity is so small that the covert
communication is practically impossible for a predefined covert communication threshold. This also indicates that
the aggregate interference is helpful to covert communication. As the density increases, the normalized secrecy
capacity increases to close 1 which implies that the received signal strength of Bob is much stronger than Willie's
signal. Indeed, as depicted in Fig. \ref{Scatteringss}, the reflected component is rather strong and easily allow
communication through reflected path, but the diffuse scattering field is very weak and mostly under the noise.

\subsubsection{The Effect of Operating Frequency}
\begin{figure}
\centering \epsfig{file=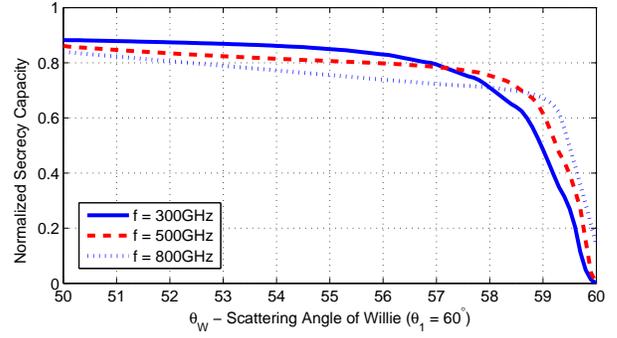, height=1.8in}
\caption{The normalized secrecy capacity $\bar{c}_s$ versus the scattering angle of Willie $\theta_W (50^{\circ}\cdots 60^{\circ})$ for different operating frequencies.
Here the incidence angle of Alice $\theta_1 = 60^{\circ}$, the surface height variation $\sigma_h=0.058$mm, and the network density $\lambda=0.01$.}\label{frequency}
\end{figure}

Next considering the effect of operating frequency $f$ on $\bar{c}_s$. Fig. \ref{frequency} shows the
comparison when different operating frequencies are taken into account. We can notice that the secrecy capacity
increases with the frequency when the scattering angle is close to the specular reflection direction, but decreases
with the frequency when the receiver angle of Willie gradually deviates from the reflection direction. This is
reasonable since the scattering always increases with the operating frequency. As discussed in \cite{Frequency}
and \cite{Diffuse_1}, when the frequency increases, it will give less energy at the reflected path, but more
energy in the diffuse scattering directions. This also implies that the normalized secrecy capacity should
decreases along with the increase of frequency.

\subsubsection{The Effect of Surface Roughness}
\begin{figure}
\centering \epsfig{file=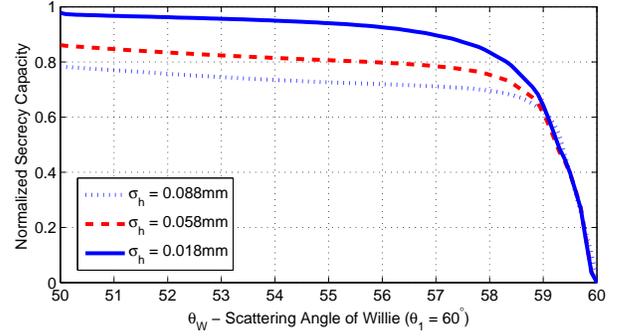, height=1.8in}
\caption{The normalized secrecy capacity $\bar{c}_s$ versus the scattering angle of Willie $\theta_W (50^{\circ}\cdots 60^{\circ})$ for different surface roughnesses $\sigma_h$.
Here the incidence angle of Alice $\theta_1 = 60^{\circ}$, the operating frequency $f=500$GHz, and the network density $\lambda=0.01$.}\label{sigma_h}
\end{figure}

The effect of surface roughness on the secrecy capacity $\bar{c}_s$ is illustrated in Fig. \ref{sigma_h}. In this
measurement, we fix the surface correlation length $l_c$, only change the standard deviation of the surface
height distribution $\sigma_h$ which gives information about the overall height variations on the surface. We
notice that the larger value of $\sigma_h$ results in lower normalized secrecy capacity. The underlying reason for
this behavior is that, for smaller value of $\sigma_h$, the surface is a more smooth surface with a purely
specular reflection component, a larger value of $\sigma_h$ would represent a relatively more rough surface
with a stronger diffuse scattering contribution \cite{Diffuse_3}.

\subsubsection{The Effect of Bob's Scattering Angle in Diffuse Scattering Covert Communications}
In practice, the reflected components are rather strong and easily allow covert communication through reflected
paths. However, Alice and Bob cannot always find a specular reflection path to perform their NLOS
communications. Normally, the signal came from the diffuse scattering field is very weak and mostly under the
noise, but in some cases it is high enough to enable NLOS communications. As an alternative, Alice and Bob may
use diffuse scattering to perform covert communications.

\begin{figure}
\centering \epsfig{file=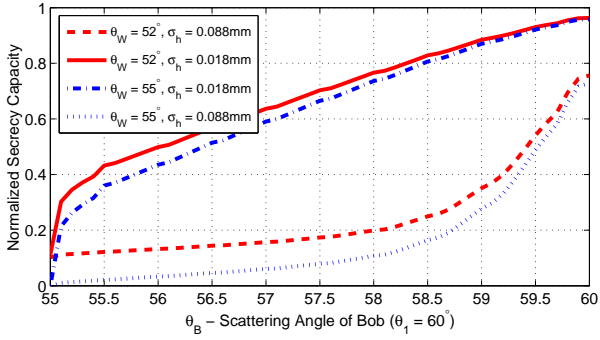, height=1.8in}
\caption{The normalized secrecy capacity $\bar{c}_s$ versus the  scattering angle of Bob $\theta_B (55^{\circ}\cdots 60^{\circ})$ for different surface roughnesses $\sigma_h$.
Here the incidence angle of Alice $\theta_1 = 60^{\circ}$, the operating frequency $f=500$GHz, and the network density $\lambda=0.01$.}\label{scattering}
\end{figure}

Fig. \ref{scattering} demonstrates the effect of Bob's scattering angle $\theta_B$ on the normalized secrecy
capacity $\bar{c}_s$. Given the incidence angle of Alice $\theta_1 = 60^{\circ}$, we fix the receiver angle of
Willie $\theta_W$ at $52^{\circ}$ and $55^{\circ}$, then calculate the value $\bar{c}_s$ at different
scattering angle of Bob $\theta_B (55^{\circ}\cdots 60^{\circ})$. The results show that, the closer Bob's
scattering direction to the specular reflection direction, the larger the value of $\bar{c}_s$. On the other hand,
a more smooth surface (with less $\sigma_h$) will have less scattering strength and therefore will have larger
$\bar{c}_s$. However, if the scattering angle deviates from the direction of reflection for several degree, the
value of $\bar{c}_s$ will decrease rapidly, especially when the surface is rougher.

\subsubsection{The Effect of Incident Angle}
\begin{figure}
\centering \epsfig{file=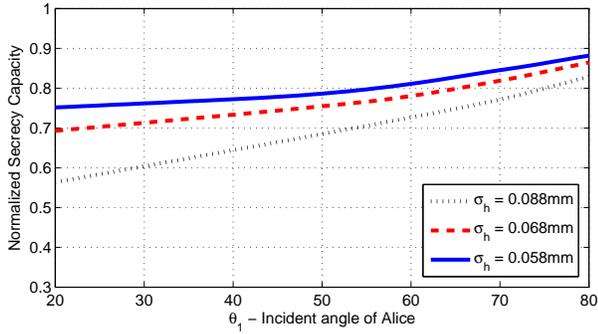, height=1.8in}
\caption{The normalized secrecy capacity $\bar{c}_s$ versus the incident angle of signal $\theta_1 (20^{\circ}\cdots 80^{\circ})$ for different surface roughnesses $\sigma_h$.
Here the scattering angle of Willie $\theta_W = \theta_1 - 5^{\circ}$, the operating frequency $f=500$GHz, and the network density $\lambda=0.01$.}\label{incident_angle}
\end{figure}

Finally we illustrate the effect of incident angle.  Fig. \ref{incident_angle} depicts the tendency of normalized
secrecy capacity $\bar{c}_s$ with the incident angle $\theta_1$. In the measurement setup, we assume Bob is
located in the reflected direction $\theta_B = \theta_1$, and Willie's receiver angle is fixed as $\theta_W =
\theta_B - 5^{\circ}$. We observe that, when the incident angle $\theta_1$ increases, the value of $\bar{c}_s$
becomes larger. However, this increase is not rapid. Besides, the more smooth the surface, the higher the
$\bar{c}_s$. This is due to the fact that a smooth surface has a stronger specular reflection component.

\subsection{Discussions}
\subsubsection{The Selection of Reflection Points}
\begin{figure}
\centering \epsfig{file=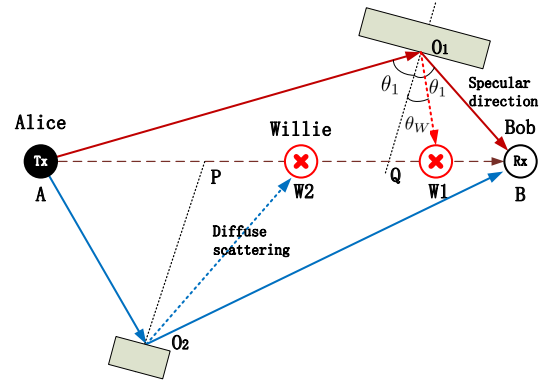, height=2in}
\caption{The selection of reflection points. $O_1$ and $O_2$ are two reflection points,
$\overline{O_1B}$ and $\overline{O_2B}$ are their specular reflection directions,
$\overline{O_1Q}$ and $\overline{O_1P}$ are the normal vectors of two scattering surfaces, respectively.
}\label{thz_model2}
\end{figure}

\begin{itemize}
  \item If Alice and Bob can find several specular reflection pathes to perform their NLOS communications,
  they should select in the first place the path whose reflection point is closest to the receiver Bob.
  In this case, Willie's detection interval will be minimized. As depicted in Fig. \ref{thz_model2}, Alice and Bob have two specular reflection pathes,
  i.e., $A\Rightarrow O_1 \Rightarrow B$ (with $O_1$ as the reflection point) and $A\Rightarrow O_2 \Rightarrow B$ (with $O_2$ as the reflection point).
  The shaded areas represent the scattering angles that Willie can eavesdrop Alice's signal.
  When we choose $O_1$ as the reflection point, Willie's eavesdropping interval is Q to B, smaller than eavesdropping interval P to B when $O_2$ is chosen as the reflection point.
   \item If there are several points with the same distance to Bob, the point with the largest incident angle $\theta_1$ will be the best choice,
   since the larger the incident angle, the higher the normalized secrecy capacity can achieve, as shown in Fig. \ref{incident_angle}.
  \item In some cases, no specular reflection path can be found. Alice and Bob have to use a scattering path to communicate.
  Because the diffuse scattering field is very weak in comparison to the specular direction,
  they should find a scattering point as close to the receiver as possible, and make the scattering direction close to the specular reflection direction.
  In certain circumstances, the diffuse scattering field may be high enough to enable NLOS communications.
  Additionally, a more rough surface with larger value of $\sigma_h$ is preferred because it provides a stronger diffuse scattering contribution,
  as illustrated in Fig. \ref{sigma_h}, Fig. \ref{scattering}, and Fig. \ref{incident_angle}.
 \end{itemize}

\subsubsection{Willie's Detection Strategy}
\begin{figure}
\centering \epsfig{file=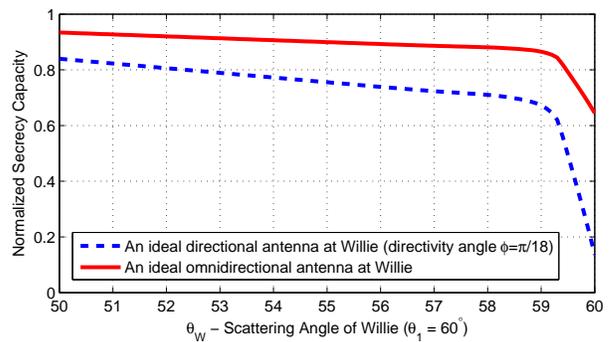, height=1.8in}
\caption{The normalized secrecy capacity $\bar{c}_s$ versus the scattering angle of Willie $\theta_W (50^{\circ}\cdots 60^{\circ})$ for different antennas of Willie.
Here the incidence angle of Alice $\theta_1 = 60^{\circ}$, the surface height variation $\sigma_h=0.058$mm, and the operating frequency $f=800$GHz.
}\label{Willie_360degree}
\end{figure}

\begin{itemize}
  \item In general, to detect the transmission between Alice and Bob, Willie should put himself in the LOS path
      between Alice and Bob,  and aim his antenna at Alice to detect their LOS transmission. After all, the LOS
      transmission is the most reasonable and effective method in THz band. However, if Willie has no
      information about the NLOS transmission channels between Alice and Bob, or does not know which
      reflected path they utilize at a particular time, he cannot put his directional antenna at the right direction.
      From this perspective, a better way is to adopt an omnidirectional antenna to find Alice's transmit signal,
      no matter which reflected path Alice and Bob use. But this method also has some drawbacks. At first the
      gain of an omnidirectional antenna is much lower than a directional antenna with a small directivity angle.
      Then, the omnidirectional antenna will experience more interference from other transmitters in the vicinity.
      Fig. \ref{Willie_360degree} also shows the normalized secrecy capacity Alice and Bob can get when Willie
      adopts an omnidirectional or directional antenna. It is important to note that the omnidirectional antenna
      has relatively lower detection capability compared with the directional antenna. But Looking from the other
      side, if Willie has no knowledge about the direction of the reflected signal, a wrong receiving direction of
      his directional antenna would be counterproductive. Therefore, how to determine the type of antenna is a
      dilemma that Willie has to be confronted with.
  \item As to Willie, the best-case scenario is that he knows the NLOS transmission path between Alice and
      Bob. However it is extremely unwise of Willie to abandon the LOS path and leave to the possible NLOS
      path. In THz band, as a result of the transmission at very high data rates, the time consumed in
      transmitting a packet can be expectedly several orders of magnitude lower than in classical wireless
      networks. Placing himself in the place between Alice and Bob, Willie can not only  block the LOS
      transmission, but also keep watch on other NLOS transmission as long as he can move close enough to Bob.
  \item The previous analysis assumes that the aim of Willie is to detect the transmission between Alice and
      Bob. In most practical scenarios, Willie only cares about whether Alice is transmitting or not. To detect the
      transmission attempt of Alice, Willie should approach Alice as close as possible, and ensure that there is no
      other node located closer to Willie than Alice. Otherwise, Willie cannot determine which one is the actual
      transmitter. But in a wireless network, some wireless nodes are probably placed on towers, trees, or
      buildings, Willie cannot get close enough as he wishes. Furthermore, wireless networks are diverse and
      complicated. If Willie is not definitely sure that there is no other transmitter in his vicinity, he cannot
      ascertain that Alice is transmitting. However, in a mobile wireless network, some mobile nodes may move
      into the detection region of Willie, and increase the uncertainty of Willie. Therefore mobile can improve the
      performance of covert communications to some extent.
\end{itemize}

\subsubsection{Covert Communication with Artificial Noise in THz Band}
If Alice has several helpers who can generate artificial noise, covert communication will become easier. Alice
simply informs these helpers to align their antennae to Willie and inject artificial noise. Due to the narrow beams,
these artificial noise will only interference Willie, not Bob.

\section{Conclusions}\label{ch_5}
In this paper, we have studied covert wireless communications with the consideration of interference uncertainty.
Prior studies on covert communications only considered the background noise uncertainty, or introduced
collaborative jammers producing artificial noise to help Alice in hiding the communication. By introducing
interference measurement uncertainty, we find that uncertainty in noise and interference experienced by Willie is
beneficial to Alice, and she can achieve undetectable communication with better performance. For AWGN
channels, if Alice want to hide communications with interference in noisy wireless networks, she can reliably and
covertly transmit $\mathcal{O}(\log_2\sqrt{n})$ bits to Bob in $n$ channel uses. Although the covert rate is
lower than the square root law, its spatial throughput is higher as $n\rightarrow\infty$. As to THz Band
networks, covert communication based on reflection or diffuse scattering is a possible information hiding way.
From the network perspective, the communications are hidden in noisy wireless networks. It is difficult for Willie
to ascertain whether a certain user is transmitting or not, and what he sees is merely a shadow wireless network.


\bibliographystyle{IEEEtran}
\bibliography{IEEEabrv,js}
\end{document}